\documentclass{elsart}
\usepackage{epsf}

\begin{document}
\begin{frontmatter}

\title{Multifractality in the stock market: \\ price increments versus 
waiting times}

\author{P.~O\'swi{\c e}cimka$^1$, J.~Kwapie\'n$^1$ and 
S.~Dro\.zd\.z$^{1,2}$}

\address{$^1$Institute of Nuclear Physics, Polish Academy of Sciences,
PL--31-342 Krak\'ow, Poland \\ 
$^2$Institute of Physics, University of Rzesz\'ow, PL--35-310 Rzesz\'ow, 
Poland}

\begin{abstract}

By applying the multifractal detrended fluctuation analysis to the
high-frequency tick-by-tick data from Deutsche B\"orse both in the price
and in the time domains, we investigate multifractal properties of the
time series of logarithmic price increments and inter-trade intervals of
time. We show that both quantities reveal multiscaling and that this
result holds across different stocks. The origin of the multifractal
character of the corresponding dynamics is, among others, the long-range
correlations in price increments and in inter-trade time intervals as well
as the non-Gaussian distributions of the fluctuations. Since the
transaction-to-transaction price increments do not strongly depend on or
are almost independent of the inter-trade waiting times, both can be
sources of the observed multifractal behaviour of the fixed-delay returns
and volatility. The results presented also allow one to evaluate the
applicability of the Multifractal Model of Asset Returns in the case of 
tick-by-tick data.
 
\end{abstract}

\end{frontmatter}

\section{Introduction}

From the perspective of the classical financial market models, behaviour
of the consecutive price and index fluctuations does not present any
significant time autocorrelations except for short time scales up to
several minutes. As regards the dynamical character of these fluctuations,
they are considered as being to a good approximation the fractional
Gaussian noise~\cite{mandelbrot68}, with very small and negligible
probability of the occurrence of non-Gaussian large jumps in the index or
the share price.  As a natural consequence, the stock market data is
expected to present only monofractal properties. However, these
widely-used models do not describe the processes underlying the evolution
of financial data with satisfactory precision. The so-called financial
stylized facts comprising, among others, the non-negligible fat tails of
log-return distributions, volatility clustering and its long-time
correlations, anomalous diffusion
etc.~\cite{plerou99,gopikrishnan99,plerou00,drozdz03} counter the
above-mentioned fundamental assumptions of market dynamics challenging
their applicability in practice. That the financial dynamics is more
complex than it is commonly assumed can also be inferred from a number of
recently-published papers discovering and exploring the multifractal
characteristics of data from the stock
markets~\cite{pasquini99,ivanova99,ausloos02,bershadskii03,matteo04}, the
forex
markets~\cite{ivanova99,matteo04,fisher97,vandewalle98,bershadskii99} and
the commodity ones~\cite{matia03}. The concept of multifractality was
developed in order to describe the scaling properties of singular measures
and functions which exhibit the presence of various distinct scaling
exponents in their different parts~\cite{halsey86,barabasi91}. Soon the
related formalism was successfully applied to characterize empirical data
in many distant fields like turbulence~\cite{muzy91,ghasghaie96}, earth
science~\cite{ashkenazy02}, genetics~\cite{peng94,buldyrev95,arneodo98},
physiology~\cite{ivanov99,blesic99,hausdorff01} and, as already mentioned,
in finance. The problem of detecting multifractality in real data is
delicate, however. There are models based on fractal processes which are
able to mimic the real multifractal evolution of markets being either
multifractal or
monofractal~\cite{mandelbrot97,lux03b,muzy01,eisler04,bouchaud00}.  
Moreover, as it has been pointed out (\cite{lux03b,nakao00}), the power of
commonly-used tests of multifractality is limited, because they cannot
effectively distinguish between the two types of fractal behaviour of the
financial (but perhaps also other) data. One important source of this
difficulty is the presence of non-Gaussian tails in the distributions of
data (e.g. truncated L\'evy~\cite{nakao00}), the fact which is ubiquitous
in finance. Thus, all conclusions drawn from multifractal analysis have to
be interpreted with care.

In the present paper we analyze data from the German stock market focusing
on their fractal properties.  We apply the multifractal detrended
fluctuation analysis which is a well-established method of detecting
scaling behaviour of signals. By exploiting the character of the
high-frequency transaction-by-transaction recordings for the most liquid
stocks belonging to DAX, we are able to inspect not only the properties of
share price fluctuations, but also the properties of time intervals
between consecutive trades (waiting times). The majority of analyses
carried out so far was devoted to time series of the returns
calculated with some fixed time delay $\Delta t$. According to the
Multifractal Model of Asset Returns introduced by Mandelbrot and
others~\cite{mandelbrot97,mandelbrot97b,lux03a,lux03b,eisler04}, the
source of multifractality in the returns is a deformation of time
$\theta(t)$, which takes place due to the fact that at the microscale the
so-called business time is ``paced'' by transactions rather than any
constant time units. In this model the increments of so-deformed time can
be directly related~\cite{plerou00,eisler04} to the fixed-$\Delta t$
volatility which depends both on the trading activity (i.e.  
$\sim\sqrt{N^{(\Delta t)}(t)}$, where $N^{(\Delta t)}$ stands for the
number of transactions in $\Delta t$) and on the variance $W^{(\Delta
t)}(t)$ of the price change in the individual transactions over $\Delta
t$. Looking deeper into the microstructure of the market, as these
quantities are equal to, respectively, the reciprocal of the average
waiting time in $\Delta t$ and to the average squared price increment in
$\Delta t$, one may ask whether, at the microscopic level, the
multifractal properties of $\theta(t)$ originate from the price behaviour,
from the waiting times fluctuations or from both of them. This is our
motivation for performing the present analysis. We find the German stocks
particularly suitable for such a study due to the fact that the moments of
transactions are recorded with high precision (0.01 s), which almost
completely eliminates falsely simultaneous transactions. Such an analysis
for the liquid stocks from NYSE, although desired, cannot be so
successfully carried out as the time resolution of the recordings is poor
(1 s) and too many transactions are forced to be simultaneous, distorting
the underlying real dynamics.

\section{Methods and data}

There are two possible procedures of analyzing multifractal properties of
a time series. The first one uses the continuous wavelet transform and
extracts scaling exponents from the wavelet transform amplitudes over all
scales~\cite{arneodo95}. This method is more computationally demanding
and, as our preliminary tests showed, the stability of results for our
data is not satisfying. Therefore, for the present study we prefer to
employ the multifractal version of the detrended fluctuation analysis
method (MF-DFA)~\cite{peng94,kantelhardt02}.

Given the time series of price values $p_s(t_s(i)), i=1,...,N_s$ of a
stock $s$ recorded at the discrete transaction moments $t_s(i)$, one may
consider two independent random processes defined by price $p_s(i)$ and
time $t_s(i)$ or, alternatively, logarithmic price increments\footnote{
Throughout this paper we intend to use the expression ``price
increments'' instead of ``returns'' to underline the fact that we study 
the price differences between consecutive trades which occur irregularly 
in time, while in literature the ``returns'' are usually associated 
with a constant time step.} $g_s(i)=\ln(p_s(i+1))-\ln(p_s(i))$ and time 
increments (waiting times) $\Delta t_s(i) = t_s(i+1)-t_s(i)$.

For the time series of the log-price increments 
$G_s:=\{g_s(i)\}_{i=1,..,N_s}$ (and analogously for the waiting-times
series $T_s:=\{t_s(i)\}_{i=1,...,N_s}$), one needs to estimate the signal 
profile 
\begin{equation}
Y(i) = \sum_{k=1}^i{(g_s(k)-<g_s>)}, \ i = 1,...,N_s
\end{equation}
where $<...>$ denotes the mean of $G_s$. $Y(i)$ is divided into $M_s$
disjoint segments of length $n$ starting from the beginning of $G_s$. For
each segment $\nu, \nu=1,...,M_s$, the local trend is to be calculated by
least-squares fitting the polynomial $P_{\nu}^{(l)}$ of order $l$ to the
data, and then the variance
\begin{equation}
F^2(\nu,n) = \frac{1}{n} \sum_{j=1}^n \{Y[(\nu-1) n+j] - 
P_{\nu}^{(l)}(j)\}^2.
\label{std.dev}
\end{equation}
In order to avoid neglecting data points at the end of $G_s$ which do not 
fall into any of the segments, the same as above is repeated for $M_s$ 
segments starting from the end of $G_s$ (i.e. finally one has $2 M_s$ 
segments total and the same number of $F^2$'s). The polynomial order $l$ 
can be equal to 1 (DFA1), 2 (DFA2), etc. The variances (\ref{std.dev}) 
have to be averaged over all the segments $\nu$ 
and finally one gets the $q$th order fluctuation function
\begin{equation}
F_q(n) = \bigg\{ \frac{1}{2 M_s} \sum_{\nu=1}^{2 M_s} [F^2(\nu,n)]^{q/2} 
\bigg\}^{1/q}, \ \ q \in \mathbf{R}.
\end{equation}
In order to determine the dependence of $F_q$ on $n$, the function 
$F_q(n)$ has to be calculated for many different segments of lengths $n$.

If the analyzed signal develops fractal properties, the fluctuation 
function reveals power-law scaling
\begin{equation}
F_q(n) \sim n^{h(q)}
\label{scaling}
\end{equation}
for large $n$. The family of the scaling exponents $h(q)$ can be then 
obtained by observing the slope of log-log plots of $F_q$ vs.~$n$.
$h(q)$ can be considered as a generalization of the Hurst exponent $H$ 
with the equivalence $H \equiv h(2)$. Now the distinction between 
monofractal and multifractal signals can be performed: if $h(q)=H$ for all 
$q$, then the signal under study is monofractal; it is multifractal 
otherwise. By the procedure, $h(q), q<0$ describe the scaling properties 
of small fluctuations in the time series, while the large ones correspond 
to $h(q), q>0$. It also holds that $h(q)$ is a decreasing function of $q$.

\begin{figure}
\hspace{0.0cm}
\epsfxsize 14cm
\epsffile{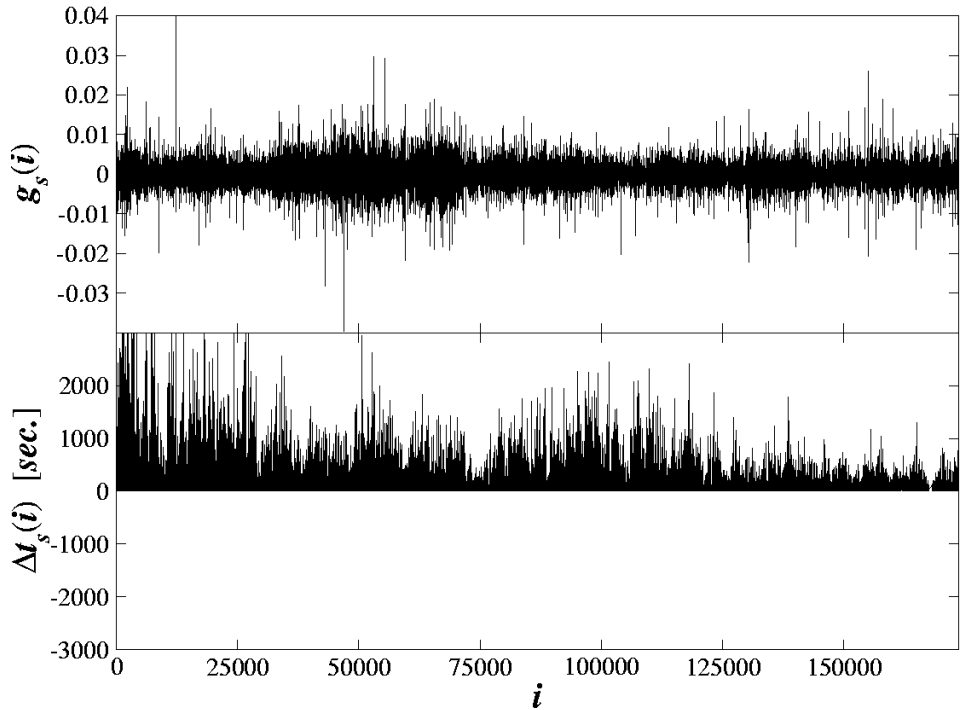}
\caption{Time course of time series of logarithmic price increments
$g_s(i)$ (upper panel) and time series of inter-trade waiting times
$\Delta t_s(i)$ (lower panel) for an exemplary stock (BMW).}
\label{signals}
\end{figure}

By knowing the spectrum of the generalized Hurst exponents, one can 
calculate the singularity strength $\alpha$ and the singularity spectrum 
$f(\alpha)$ using the following relations (e.g.~\cite{kantelhardt02}):
\begin{equation}
\alpha=h(q)+q h'(q) \hspace{1.0cm} f(\alpha)=q [\alpha-h(q)] + 1,
\label{singularity}
\end{equation}
where $h'(q)$ stands for the derivative of $h(q)$ with respect to $q$.

Our analysis was performed on the time series of tick-by-tick recordings
for the 30 DAX stocks comprising the over-two-years-long interval between
Nov 28, 1997 and Dec 31, 1999. The time series were approx. 250,000 points
long on average (almost 500 transactions daily) with the minimal length of
63,000 (Karstadt Quelle, KAR) and the maximal one of 588,000
(Daimler-Chrysler, DCX). Firstly, we removed all the overnight price
increments, because they correspond to extremely long inter-trade
intervals and larger-than-usual price jumps (such effects were identified
to introduce distortion of the financial scaling~\cite{gorski03}); we do
not remove zero increments from $G_s$ and zero waiting times from $T_s$
because the corresponding zero-intervals are so short that they do not
alter the results. The so-preprocessed data was subject to the MF-DFA
analysis with the polynomials $P^{(3)}$ (MF-DFA(3)), owing to the fact
that for the present data they optimally extract the fluctuations whose
scaling is to be analyzed (Eq.~\ref{scaling}).

\section{Results}

\begin{figure}
\hspace{0.0cm}
\epsfxsize 14cm
\epsffile{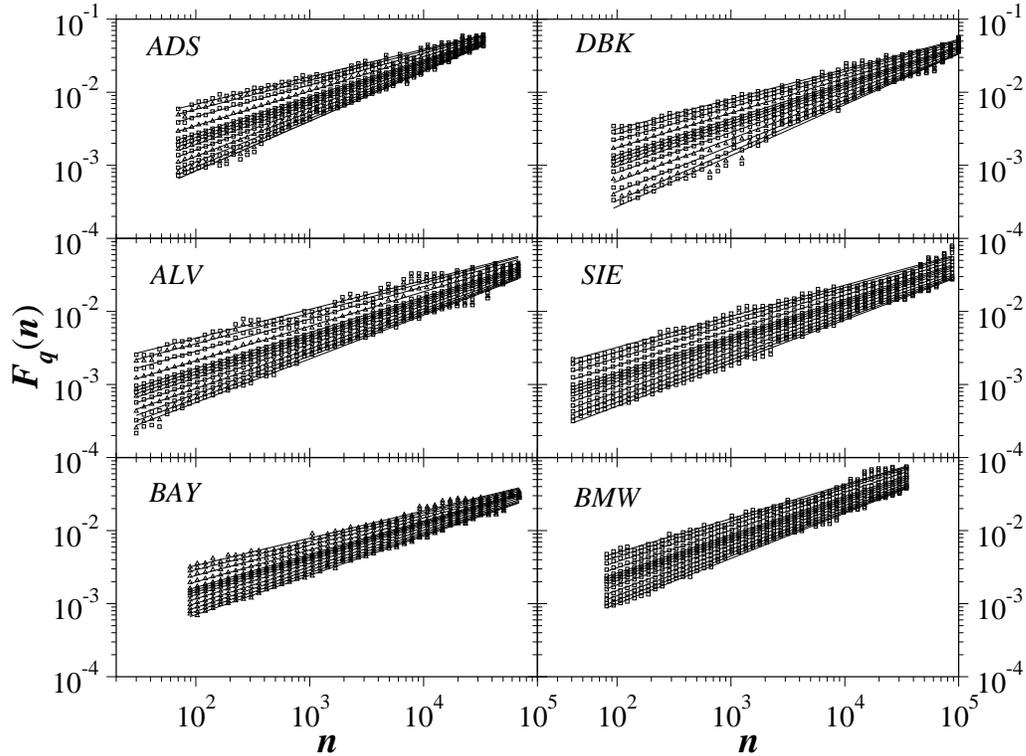}
\caption{Log-log plots of the $q$-th order fluctuation $F_q$ for time
series of price increments as a function of segment size $n$ for 
different values of $q$ between -10 (bottom line) and 10 (top line). Six 
stocks are shown ordered according to the slope spread between the extreme 
values of $q$ (left to right and top to bottom). In each panel only a 
region of the scaling regime is shown and a non-scaling part for small $n$ 
has been cut off. The largest value of $n$ depends on the time series 
length being different for each stock. Scaling regions allow one to 
estimate $h(q)$ according to Eq.~\ref{scaling}.}
\label{retn-slope}
\end{figure}

Figure 1 shows the time course of $G_s$ (upper panel) and $T_s$ (lower 
panel) for the complete interval under study. There is a clear difference 
between the series, because while the distribution of the former is 
symmetric around zero, the latter cannot assume negative values and, 
hence, its distribution is skewed. 

\begin{figure}
\hspace{0.0cm}
\epsfxsize 14cm
\epsffile{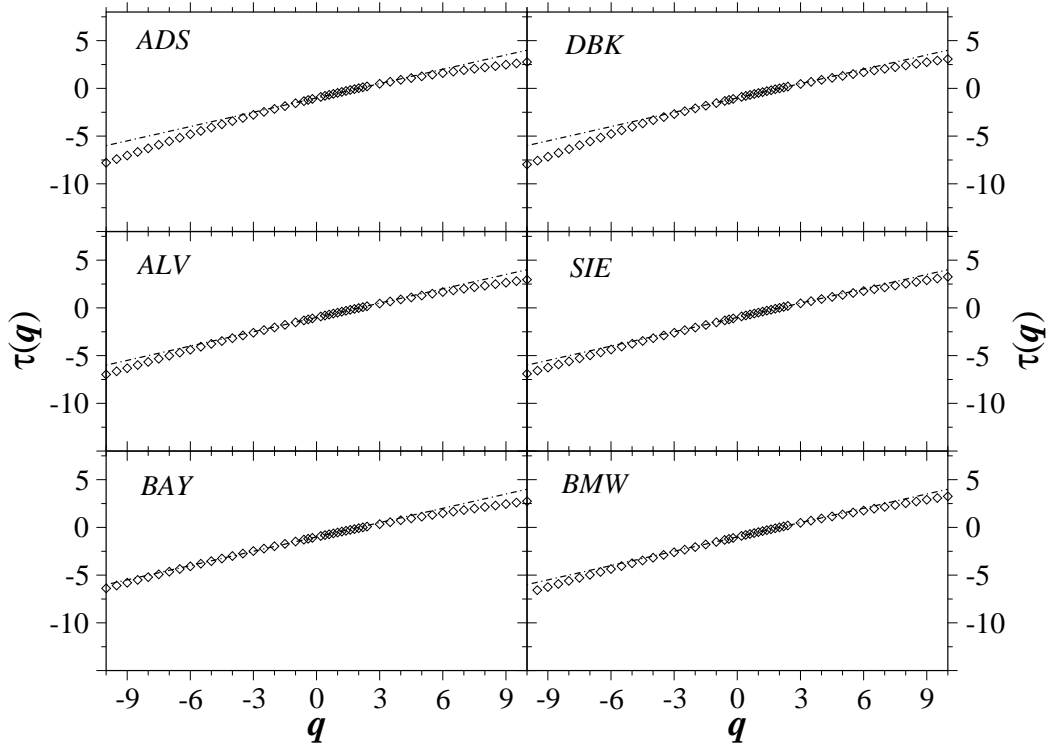}
\caption{Multifractal spectra expressed by $\tau(q)=q h(q)-1$ for 
price increments; dot-dashed line in each panel presents a monofractal 
Gaussian noise regime. A nonlinear behaviour of $\tau(q)$ can be 
considered a manifestation of multiscaling. The same stocks as in Fig.~2 
are shown.}   
\label{retn-multifractal}
\end{figure}   

First we shall present results of our study of the price increments data.  
Following Eq.~(\ref{scaling}), for each stock we created
double-logarithmic plots of $F_q(n)$ for the various segment lengths in
the range $30 \le n \le N_s / 5$ data points and for various choices of
$q$. It is widely assumed that the amplitude of the price fluctuations
scale according to the inverse cubic power law, which in principle implies
infinite moments for $q \ge 3$. In real data the large-order moments are
therefore inevitably affected by the finite-size effects. However, we
decided to include the values of $q$ as high as 10 purely to improve
readability of the plots of the multifractal and singularity spectra and
to make any comparison of the spectra's widths easier; this step does not
reduce the significance of our results. The plots of $F_q(n)$ for six
representative stocks from different market sectors are collected in
Fig.~2 and ordered according to the decreasing slope spread between
$q=-10$ and 10 (with the step of 0.2 for small $|q|$'s and of 0.5 for
larger ones). In order to indicate the range of $n$'s used for fitting the
exponents $h(q)$, we present the scaling regions only and cut off the
regions characterized by the lack of scaling (this happens for small
$n$'s, probably due to the existence of long periods of constant price of
a stock). The largest difference between the slopes is observed for Adidas
Solomon (ADS, top left panel) and the smallest one for BMW (bottom right
panel). The plots for different companies show noticeable differences in
the range of $n$ with the scaling behaviour; the widest range of over
three decades is observed for moderate values of $q$ for ALV (Allianz) and
SIE (Siemens). In each case, for extremely small and extremely large $n$'s
the scaling breaks down as expected from statistical
considerations~\cite{kantelhardt02}. In order to better visualize the
scaling character of the data, in Fig.~3, we show the corresponding
multifractal spectra. Instead of $h(q)$, we display its function
$\tau(q)$, defined by the relation $\tau(q)=q h(q)-1$. Monofractal signals
($h(q)=$const) are associated with a linear plot $\tau(q)$, while
multifractal ones possess the spectra nonlinear in $q$. Keeping in mind
the already-mentioned limitations of the method (\cite{lux03a}), our
calculations indicate that the time series of price increments for all
companies can be of the multifractal nature.  Consistently with the
log-log plots on the left, the highest nonlinearity of the spectrum and
the strongest multifractality are attributes of ADS and DBK (Deutsche
Bank), and the smallest nonlinearity and the weakest multifractal
character correspond to BMW and BAY (Bayer). The nonlinearity of $\tau(q)$
is confined to the central range of $q$'s around $q=0$ and for larger
values of $|q|$ the behaviour of $\tau(q)$ is almost linear (due to the
finite size of the sample).

\begin{figure}
\hspace{0.0cm}
\epsfxsize 14cm
\epsffile{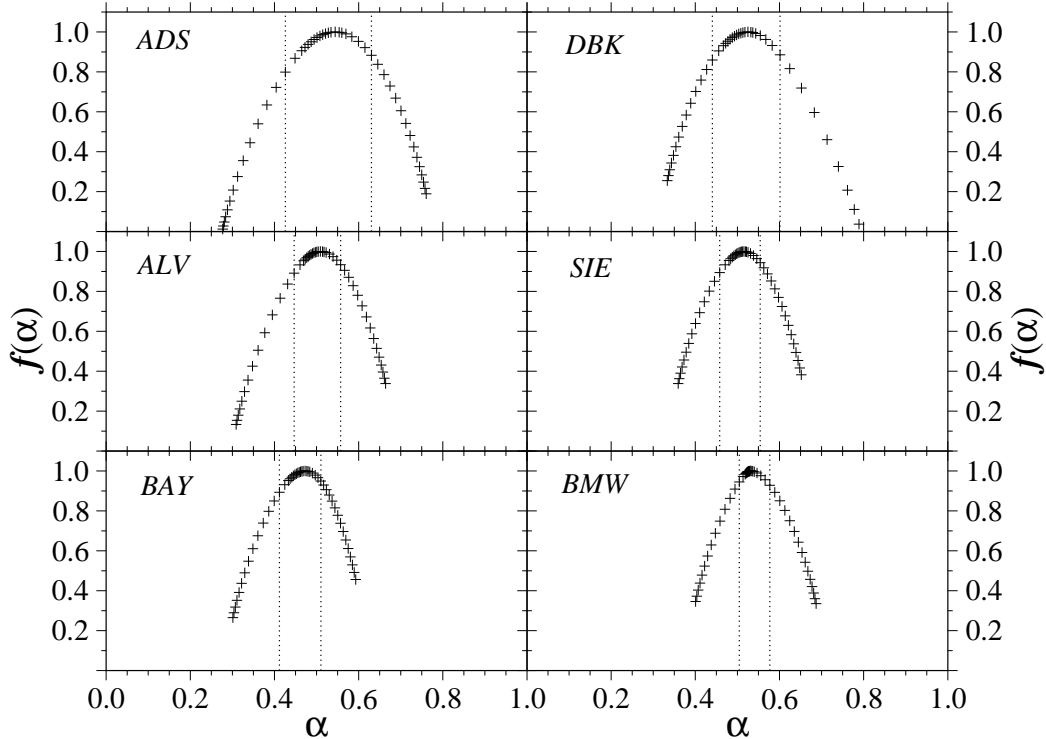}
\caption{Singularity spectra $f({\alpha})$ vs. $\alpha$ for price 
increments and for the same stocks as in Fig.~2. The maximum value of each 
spectrum gives the most common singularity strength $\alpha_0$. The 
vertical dotted lines denote $q = -3.0$ (right) and $q=3.0$ (left).}
\label{retn-singularity}
\end{figure}

\begin{figure}
\hspace{0.0cm}
\epsfxsize 14cm
\epsffile{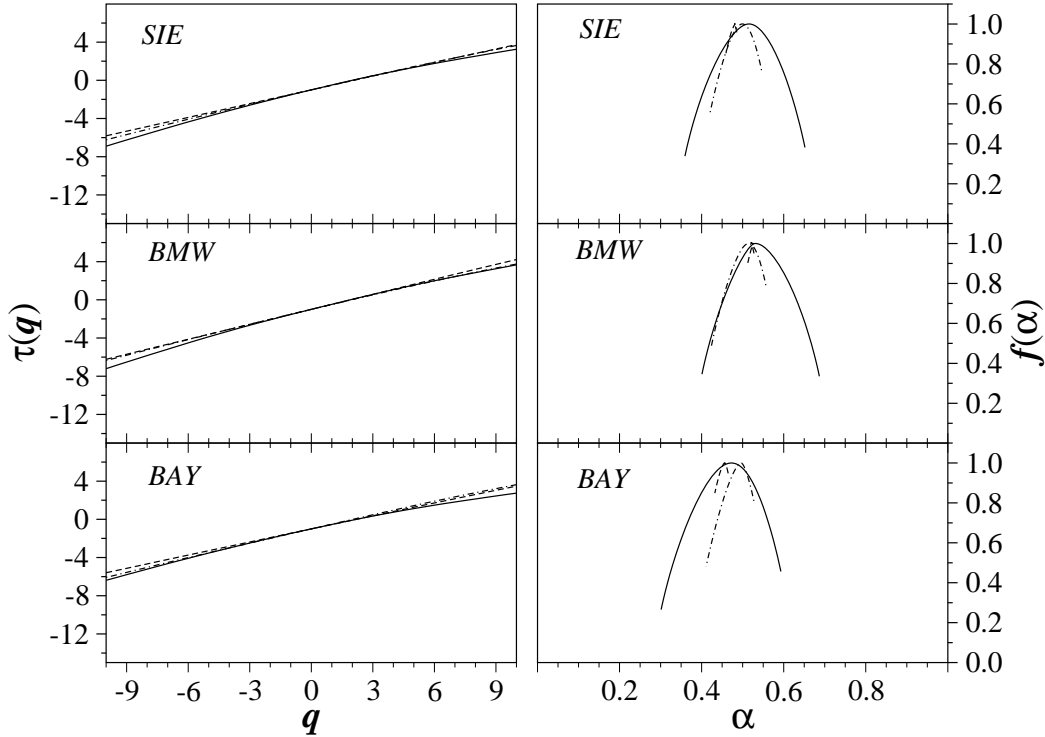}
\caption{Comparison of the original and randomized price increments: 
original (solid), reshuffled (dashed) and surrogate (dotted) time series.
Multifractal spectra $\tau(q)$ (left column) and singularity spectra
$f(\alpha)$ (right column) are presented for three different stocks.}
\label{time-slope}
\end{figure}

The multifractal nature of the data can also be expressed in a different
manner, i.e. by plotting the singularity spectra $f(\alpha)$
(Eq.~\ref{singularity}). It is a more plausible method because here one
can easily assess the variety of scaling behaviour in the data. Figure 4
displays such spectra for the same stocks as in Figs.~2 and 3, with their
presentation order being preserved. As before, the richest multifractality
(the widest $f(\alpha)$ curve) is visible for ADS
($\Delta\alpha:=\alpha_{q=-10}-\alpha_{q=10}\simeq 0.5$) and DBK
($\Delta\alpha\simeq 0.45$), the poorest one for BAY and BMW
($\Delta\alpha\simeq 0.3$). The maxima of $f(\alpha)$ are typically placed
in a close vicinity of $\alpha=0.5$ indicating no significant
autocorrelations exist. It is worth mentioning, that the singularity
spectra for the price changes between two consecutive transactions closely
resemble their counterparts for the fixed-$\Delta t$ log-returns. We do
not show the corresponding results graphically, but the widths of the
$f(\alpha)$ spectra are similar in both cases.

The multifractal character of price fluctuations can originate from the
existence of the long-range correlations in the price increments (via
volatility) as well as from their non-Gaussian
distributions~\cite{nakao00}.  The possible influence of each of these
factors can be detected by a proper modification of the data. The
long-range autocorrelations can be completely erased by randomly
reshuffling the original time series and the non-Gaussianity of the
distributions can be weakened by creating the phase-randomized
surrogates~\cite{theiler92}. In the latter case we exploit the fact that
the price increments distributions are unstable in the sense of L\'evy,
which leads to their convergence to a Gaussian under the discrete Fourier
transforms. Figure 5 shows three examples of $\tau(q)$ (left column) and
$f(\alpha)$ (right column) spectra for the original (solid), reshuffled
(dot-dashed) and phase-randomized (dashed)  data. Both the widths of the
$f(\alpha)$ spectra in each case are much smaller and the nonlinearity of
$\tau(q)$'s is much weaker for the modified signals than for the original
ones. This behaviour of the reshuffled signals confirms that the
persistent autocorrelations play an important role in multiscaling of the
price increments. However, the spectra for the surrogates are typically
much narrower than for the reshuffled data which can be interpreted as an
evidence of the influence of extremely large non-Gaussian events on the
fractal properties of the signals. Keeping in mind the difficulties in
precisely calculating the scaling exponents $h(q)$ for finite time series,
one may interpret the narrow curves for the modified signals in Fig.~5 as
the manifestation of their relatively, but not precisely, monofractal
character.

To this end, we concentrated on multifractal properties of the time series
of logarithmic price increments. Our results go in parallel with earlier 
analyses of other groups, which managed to show multifractality in the 
stock market data~\cite{ausloos02,vandewalle98,matia03}. Once again, we 
stress that what distinguishes our approach is that we did not sample the 
data with fixed time interval, but rather look into the tick-by-tick 
data and constructed the time series of price increments for variable 
inter-trade time intervals (thus unfolding time). Now we shall go back to 
the original time axis, and study properties of the inter-trade time 
intervals forming the series $T_s$.

\begin{figure}
\hspace{0.0cm}
\epsfxsize 14cm
\epsffile{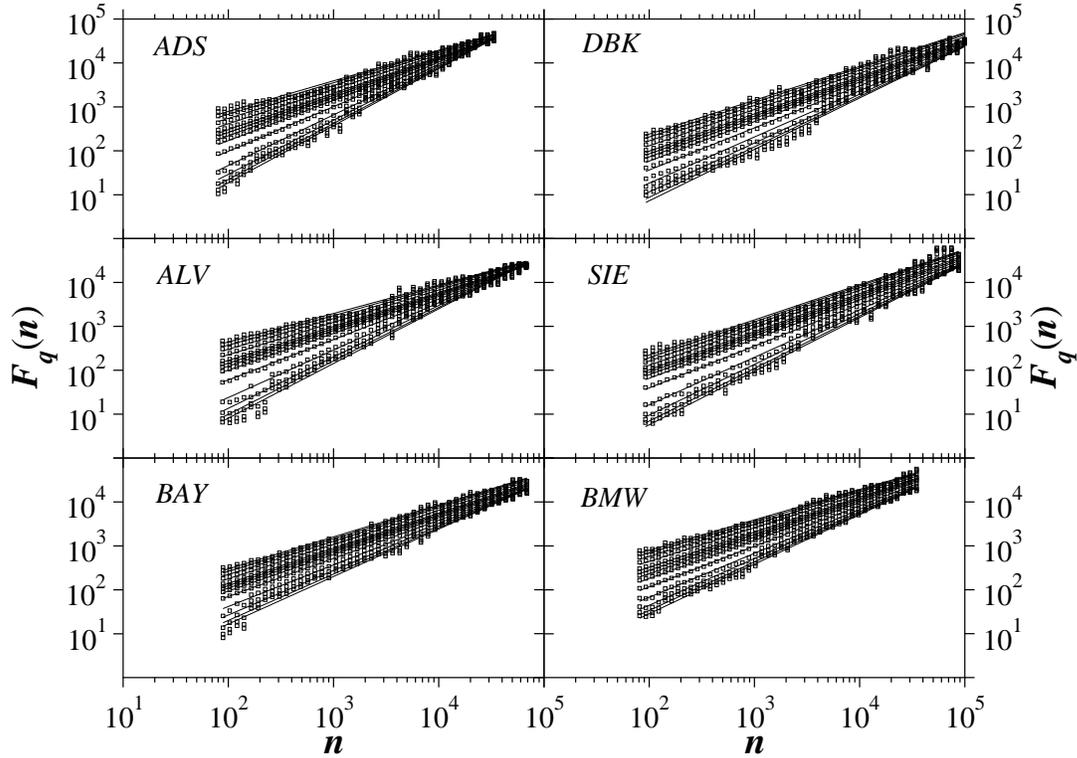}
\caption{Log-log plots of $F_q$ for waiting times as a function
of $n$ ($-10 \le q \le 10$). This figure is the analog of Fig.~2 for 
$G_s$; the same six stocks are shown.}
\label{time-slope}
\end{figure}

\begin{figure}
\hspace{0.0cm}
\epsfxsize 14cm
\epsffile{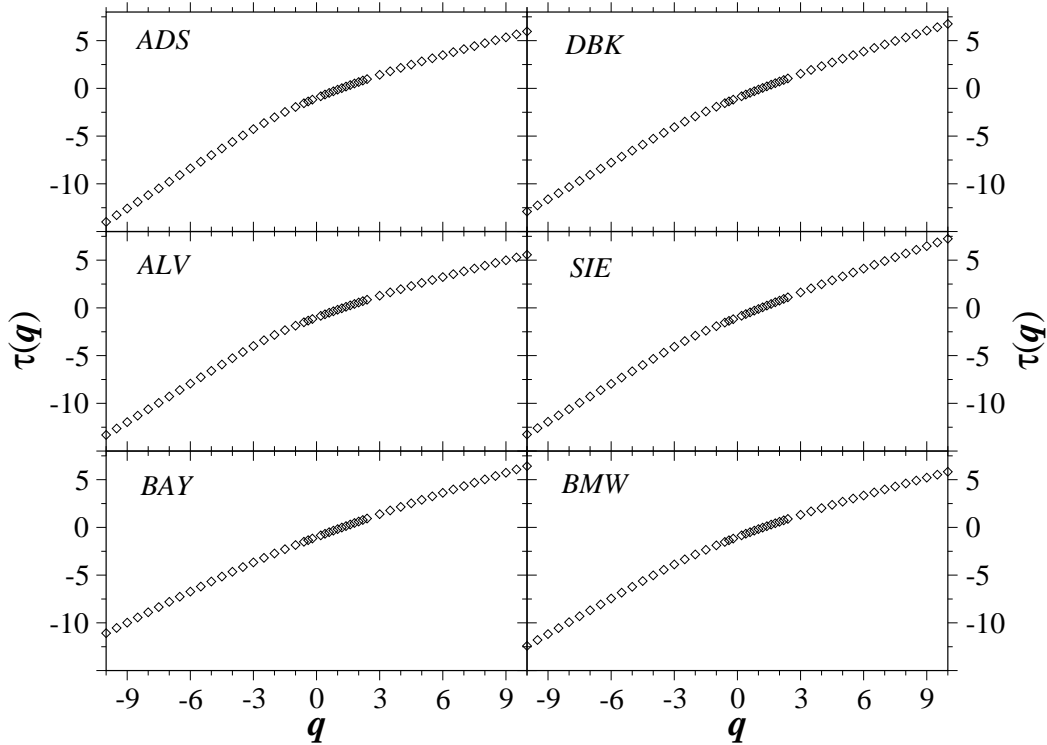}
\caption{Multifractal spectra $\tau(q)=q h(q)-1$ for waiting times. A 
nonlinear behaviour of $\tau(q)$ can be considered a manifestation of 
multiscaling.}
\label{time-multi}
\end{figure}

Figures 6, 7, and 8 are fully analogous to Figs.~2, 3, and 4,
respectively, but for the waiting times instead of the price increments.
As before, we observe significant difference between the slopes for the
most negative and the most positive $q$'s with this strength being
different among the stocks. The scaling in Fig.~6 is good for small and
moderate values of $|q|$, but is poorer for large $|q|$'s for the majority
of companies (the effect of finite size and noise). The $\tau(q)$ spectra
in Fig.~7 show strong nonlinearity, and although it is confined to middle
range of $q$'s, this nonlinearity is stronger than for the price
increments. By comparing Figs.~2-4 and 6-8 we see that there is no
systematic evidence of the relations between the properties of $G_s$ and
$T_s$ for the given companies. Fig.~8 is a counterpart of Fig.~4 and
presents the singularity spectra for the time series of the waiting times.
All the spectra can be interpreted as multifractal. The main differences
are the larger widths of $f(\alpha)$'s, their asymmetry and the positions
of maxima at approx. $\alpha \ge 0.8$ instead of 0.5 for price
fluctuations.  The widest spectra correspond to ADS and ALV, the
significantly narrower ones to BAY, in different order than for $G_s$. The
shift of the maxima can be related to long-range correlations between
consecutive waiting times, leading to a strong persistence (see also
Fig.~9).  However, despite the fact that both the spectra for $G_s$ and
for $T_s$ are multifractal, they cannot be directly compared to each
other. The crucial factor here is that they represent processes of
different character: signed and unsigned, respectively. There exists a
method of rescaling the $f(\alpha)$ spectra of the unsigned (or signed)
process (as it was proved in~\cite{mandelbrot97b}), but it requires that
the underlying processes $X(t), Y(t)$ are strongly interrelated through
the fractional Brownian motion $X(t)=B_H[Y(t)]$ with some Hurst exponent
$H$ (then simply $f_X(\alpha)=f_Y(\alpha/H)$). However, in the context of
our data, this assumption seems to be violated: $G_s$ and $T_s$ can in
principle be independent or, at least, might not be mutually related in so
simple way.  But, nevertheless, the fact that the singularity spectra for
$T_s$ are wider than their counterparts for $G_s$ and shifted towards
larger $\alpha$'s is still in the spirit of the above-mentioned relation
between the signed $f_X$ and unsigned $f_Y$ processes.

Our results suggest that as regards the original time series of price 
increments $p_s(t_s(i))$ without any unfolding of the time axis, in that 
case we can well deal with a fully two-dimensional multifractal process, 
multiscaling both in the argument $t_s(i)$ and in the value $p_s(t_s(i))$.
This may suggest that both the price increments and the waiting times 
contribute to the multifractal properties of the time deformation 
$\theta(t)$ and of the fixed-$\Delta t$ log-returns.

\begin{figure}
\hspace{0.0cm}
\epsfxsize 14cm
\epsffile{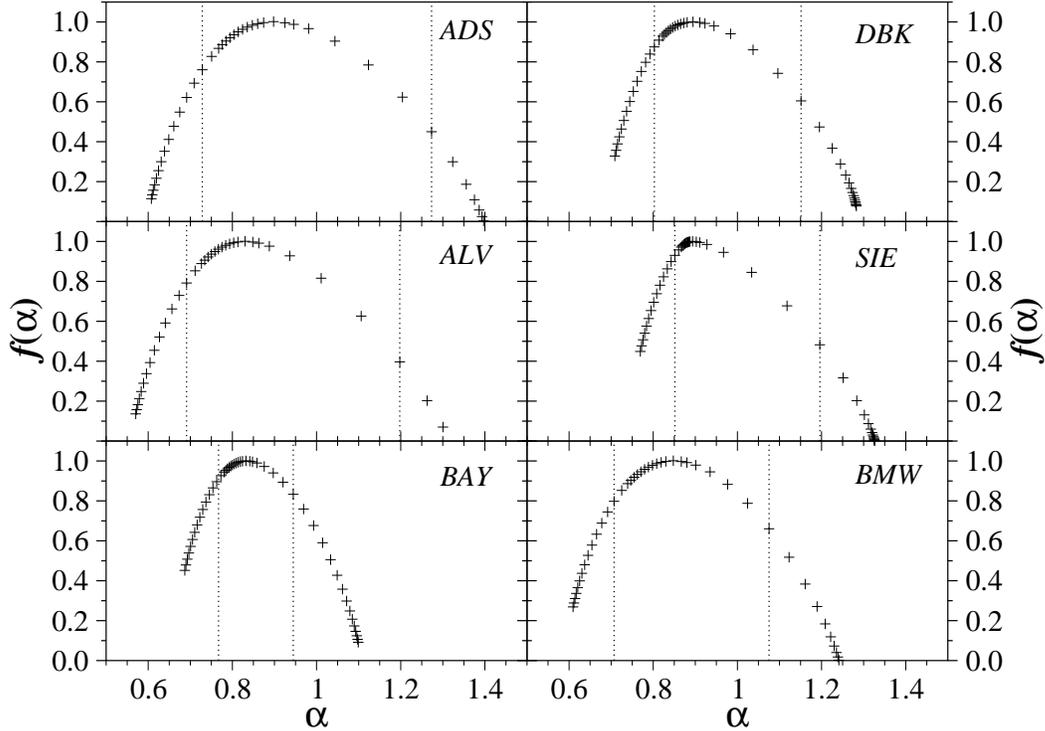}
\caption{Singularity spectra $f({\alpha})$ for waiting times; the maximum 
value of each spectrum gives the most common singularity strength 
$\alpha_0$. The vertical dotted lines denote $q = -3.0$ (right) and 
$q=3.0$ (left).}
\label{time-singularity}
\end{figure}

\begin{figure}
\hspace{0.0cm}
\epsfxsize 14cm
\epsffile{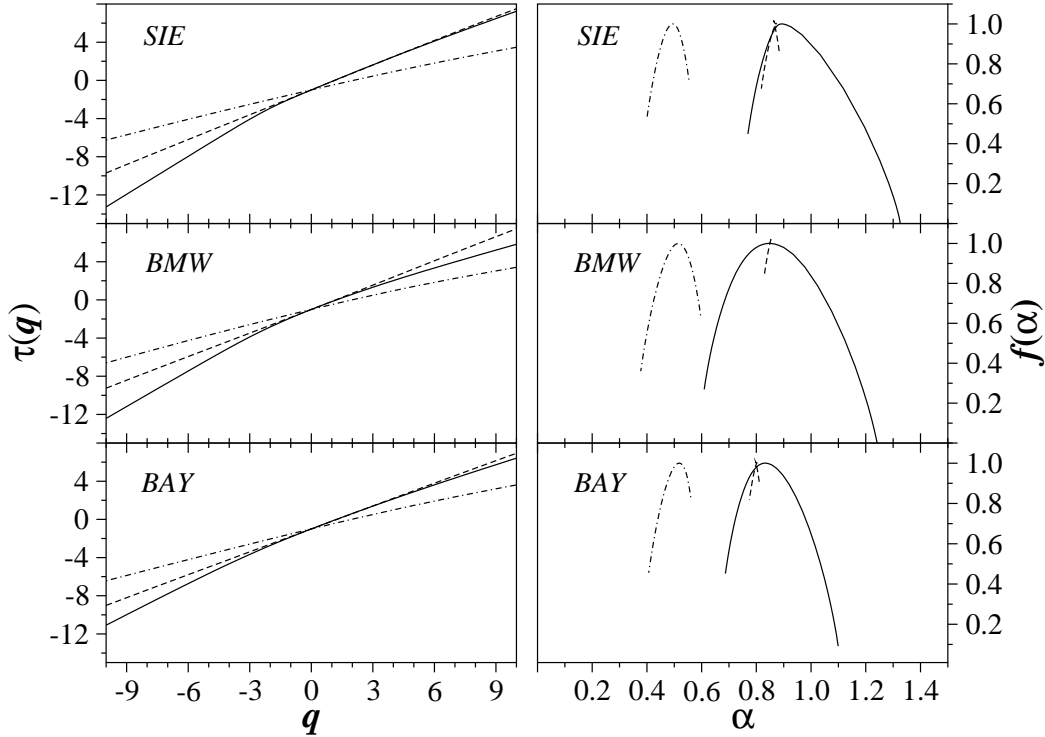}
\caption{Comparison of the original and randomized series of waiting 
times: original (solid), reshuffled (dashed) and surrogate (dotted) time 
series. Multifractal spectra $\tau(q)$ (left column) and singularity 
spectra $f(\alpha)$ (right column) are presented for three different 
stocks. Note the maximum positions for the surrogates with the linear 
correlations being preserved.}
\label{time-surrogates}
\end{figure}

Finally, Figure 9 shows a comparison of the spectra $f(\alpha)$ for the
original $T_s$ (solid), the reshuffled (dot-dashed) and the surrogate time
series (dashed). Contrary to Fig.~5, here the randomized data displays
completely different behaviour, being not only narrowed but also
systematically shifted towards $\alpha=0.5$. This is not surprising,
however: reshuffling removes the strong autocorrelation of the inter-trade 
time intervals completely. It can also be noted that the spectra for the 
surrogate signals preserve the positions of the maxima at 0.8; this is a 
natural consequence of the strong linear correlations still present in the 
surrogates.

\section{Conclusions}

We study the multifractal properties of the most liquid stocks from the
German stock market. Our original data consisting of the recordings of
time and price at which all the transactions took place, allowed us to
separate the complete process of the stock trading into its pure price and
pure time components. Since both components can contribute to the
fixed-$\Delta t$ volatility and returns, studying the properties of these
components can help to identify the microscopic sources of the observed
multifractality of the fixed-delay returns. We show that both the signals
for the transaction-to-transaction price increments and for the
inter-trade waiting times exhibit the characteristics that can be
interpreted in terms of multifractality. Its degree expressed by e.g. the
widths of the singularity spectra $f(\alpha)$ varies across different
stocks but these properties are entirely company-specific and are not
related to industry sectors, company size, average transaction frequency
or any other characteristics of this type. The multifractal properties of
$G_s$ and $T_s$ are of different nature; though on a qualitative level the
corresponding spectra can be related. The relevant
relation~\cite{mandelbrot97b} between the $f(\alpha)$ spectra for the
returns and for the multifractal time $\theta(t)$ does not however apply
fully quantitatively. This is because the c.d.f. of the microscopic price
increments and the c.d.f. of the inter-trade time intervals are not
related through the Brownian motion as required in~\cite{mandelbrot97b}.
If the price and the time components of the trading are independent or
they at most weakly depend on each other, the compelling next step in this
kind of analysis is to perform a fully 2-D multifractal approach in order
to link the fractal nature of the price and the time increments into one
unified frame.


\begin{thebibliography}{99}

\bibitem{mandelbrot68} B.B.~Mandelbrot and J.W.~van Ness, SIAM Review {\bf 
10} (1968) 422-437

\bibitem{plerou99} V.~Plerou, P.~Gopikrishnan, L.A.N.~Amaral, M.~Meyer and
H.E.~Stanley, Phys.~Rev.~E {\bf 60} (1999) 6519-6529

\bibitem{gopikrishnan99} P.~Gopikrishnan, V.~Plerou, L.A.N.~Amaral,
M.~Meyer and H.E.~Stanley, Phys.~Rev.~E {\bf 60} (1999) 5305-5316

\bibitem{plerou00} V.~Plerou, P.~Gopikrishnan, L.A.N.~Amaral, X.~Gabaix
and H.E.~Stanley, Phys.~Rev.~E {\bf 62} (2000) R3023-R3026

\bibitem{drozdz03} S.~Dro\.zd\.z, J.~Kwapie\'n, F.~Gruemmer, F.~Ruf and
J.~Speth, Acta Phys.~Pol.~B {\bf 34} (2003) 4293-4305

\bibitem{pasquini99} M.~Pasquini and M.~Serva, Economics Letters {\bf 65}
(1999) 275-279

\bibitem{ivanova99} K.~Ivanova and M.~Ausloos, Physica A {\bf 265} (1999)
279-291

\bibitem{ausloos02} M.~Ausloos and K.~Ivanova, cond-mat/0108394 (2002)

\bibitem{bershadskii03} A.~Bershadskii, Physica A {\bf 317} (2003) 591-596

\bibitem{matteo04} T.~Di Matteo, T.~Aste and M.M.~Dacorogna,
cond-mat/0403681 (2004)

\bibitem{fisher97} A.~Fisher, L.~Calvet and B.~Mandelbrot, {\it   
Multifractality of Deutschemark / US Dollar Exchange Rates}, Cowles
Foundation Discussion Paper 1166 (1997)

\bibitem{vandewalle98} N.~Vandewalle and M.~Ausloos, Eur.~Phys.~J.~B {\bf
4} (1998) 257-261

\bibitem{bershadskii99} A.~Bershadskii, Eur.~Phys.~J.~B {\bf 11} (1999) 
361-364

\bibitem{matia03} K.~Matia, Y.~Ashkenazy and H.E.~Stanley,
Europhys.~Lett.~{\bf 61} (2003) 422-428

\bibitem{mandelbrot97} B.B.~Mandelbrot, {\it Fractal and Scaling in
Finance: Discontinuity, Concentration, Risk}, Springer Verlag (New York,
1997)

\bibitem{mandelbrot97b} L.~Calvet, A.~Fisher, B.B.~Mandelbrot, {\it Large 
Deviations and the Distribution of Price Changes}, Cowles Foundation 
Discussion Paper 1165 (1997)

\bibitem{lux03a} T.~Lux, {\it The Multi-Fractal Model of Asset Returns:
Its Estimation via GMM and Its Use for Volatility Forecasting}, Univ.~of
Kiel, Working Paper (2003)

\bibitem{lux03b} T.~Lux, {\it Detecting multi-fractal properties in asset
returns: The failure of the `scaling estimator'},  Univ.~of Kiel,
Working Paper (2003)

\bibitem{halsey86} T.C.~Halsey, M.H.~Jensen, L.P.~Kadanoff, I.~Procaccia 
and B.I.~Shraiman, Phys.~Rev.~A {\bf 33} (1986) 1141-1151

\bibitem{barabasi91} A.-L. Barab\'asi and T.~Vicsek, Phys.~Rev.~A {\bf 44}
(1991) 2730-2733

\bibitem{muzy91} J.F.~Muzy, E.~Bacry and A.~Arneodo, Phys.~Rev.~Lett.~{\bf
67} (1991) 3515-3518

\bibitem{ghasghaie96} S.~Ghasghaie, W.~Breymann, J.~Peinke, P.~Talkner and
Y.~Dodge, Nature {\bf 381} (1996) 767-770

\bibitem{ashkenazy02} Y.~Ashkenazy, D.R.~Baker, H.~Gildor and S.~Havlin,
Geophys.~Res.~Lett.~{\bf 30} (2003) 2146

\bibitem{peng94} C.-K.~Peng, S.V.~Buldyrev, S.~Havlin, M.~Simons, 
H.E.~Stanley, A.L.~Goldberger, Phys.~Rev.~E {\bf 49} (1994) 1685-1689

\bibitem{buldyrev95} S.V.~Buldyrev, A.L.~Goldberger, S.~Havlin, 
R.N.~Mantegna, M.E.~Matsa, C.-K.~Peng, M.~Simons and H.E.~Stanley, 
Phys.~Rev.~A {\bf 51} (1995) 5084-5091

\bibitem{arneodo98} A.~Arneodo, B.~Audit, E.~Bacry, S.~Manneville, 
J.F.~Muzy and S.G.~Roux, Physica A {\bf 254} (1998) 24-45

\bibitem{ivanov99} P.Ch.~Ivanov, L.A.N.~Amaral, A.L.~Goldberger,
Sh.~Havlin, M.G.~Rosenblum, Z.R.~Struzik and H.E.~Stanley, Nature {\bf
399} (1999)  461-46

\bibitem{blesic99} S.~Blesic, S.~Milosevic, D.~Stratimirovic and
M.~Ljubisavljevic, Physica A {\bf 268} (1999) 275-282

\bibitem{hausdorff01} J.M.~Hausdorff, Y.~Ashkenazy, C.-K.~Peng,
P.Ch.~Ivanov, H.E.~Stanley and A.L.~Goldberger, Physica A {\bf 302} (2001)
138-147

\bibitem{muzy01} J.-F.~Muzy, D.~Sornette, J.~Delour and A.~Arneodo,
Quant.~Finance {\bf 1} (2001) 131-148

\bibitem{eisler04} Z.~Eisler and J.~Kert\'esz, cond-mat/0403767 (2004)

\bibitem{bouchaud00} J.-P.~Bouchaud, M.~Potters and M.~Meyer,
Eur.~Phys.~J.~B {\bf 13} (2000) 595-599

\bibitem{nakao00} H.~Nakao, Phys.~Lett.~A {\bf 266} (2000) 282-289

\bibitem{arneodo95} A.~Arneodo, E.~Bacry and J.F.~Muzy, Physica A {\bf
213} (1995) 232-275

\bibitem{kantelhardt02} J.W.~Kantelhardt, S.A.~Zschiegner,
E.~Koscielny-Bunde, A.~Bunde, Sh.~Havlin and H.E.~Stanley, Physica A {\bf
316} (2002) 87-114

\bibitem{gorski03} A.Z.~G\'orski, S.~Dro\.zd\.z and J.~Speth, Physica A  
{\bf 316} (2002) 496-510

\bibitem{theiler92} J.~Theiler, S.~Eubank, A.~Longtin, B.~Galdrikian,
J.D.~Farmer, Physica D {\bf 58} (1992) 77-94

\end{thebibliography}
\end{document}